\documentclass[aps,jltp,twocolumn,10pt,superscriptaddress,showpacs]{revtex4-1}

\usepackage{graphicx}
\usepackage[T1]{fontenc}
\usepackage[]{amsfonts}
\usepackage[]{amsmath}
\usepackage[]{rotating}
\usepackage[dvipsnames]{color}
\usepackage[]{float}
\usepackage[]{fancyhdr}
\usepackage[]{booktabs}
\usepackage[]{epsfig}
\usepackage{appendix}
\usepackage{amssymb}
\usepackage{psfrag}
\usepackage{setspace}

\newcommand{\ket}[1]{|#1\rangle}
\newcommand{\bra}[1]{\langle#1|}
\newcommand{\be}{\begin{equation}}
\newcommand{\ee}{\end{equation}}
\newcommand{\ba}{\begin{eqnarray}}
\newcommand{\ea}{\end{eqnarray}}

\begin{document}

\title{Highly controllable qubit-bath coupling based on a sequence of resonators}

\author{Philip J. Jones}
\affiliation{QCD Labs, COMP Centre of Excellence, Department of Applied Physics, Aalto University, P.O. Box 13500, FI-00076 Aalto, Finland}
\author{Juha Salmilehto}
\email{juha.salmilehto@aalto.fi}
\affiliation{QCD Labs, COMP Centre of Excellence, Department of Applied Physics, Aalto University, P.O. Box 13500, FI-00076 Aalto, Finland}
\author{Mikko M\"{o}tt\"{o}nen}
\affiliation{QCD Labs, COMP Centre of Excellence, Department of Applied Physics, Aalto University, P.O. Box 13500, FI-00076 Aalto, Finland}
\affiliation{Low Temperature Laboratory (OVLL), Aalto University, P.O. Box 13500, FI-00076 Aalto, Finland}

\pacs{03.67.Lx, 42.50.Pq, 85.25.Am}
\keywords{quantum bit; open quantum system; environment engineering; circuit quantum electrodynamics}

\begin{abstract}

Combating the detrimental effects of noise remains a major challenge in realizing a scalable quantum computer. To help to address this challenge, we introduce a model realizing a controllable qubit-bath coupling using a sequence of LC resonators. The model establishes a strong coupling to a low-temperature environment which enables us to lower the effective qubit temperature making ground state initialization more efficient. The operating principle is similar to that of a recently proposed coplanar-waveguide cavity (CPW) system, for which our work introduces a complementary and convenient experimental realization. The lumped-element model utilized here provides an easily accessible theoretical description. We present analytical solutions for some experimentally feasible parameter regimes and study the control mechanism. Finally, we introduce a mapping between our model and the recent CPW system.

\end{abstract}

\maketitle

\section{Introduction}

Achieving a balance between the seemingly competing demands of qubit addressibility and coherence time remains a fundamental obstacle on the path to engineering a large scale quantum computer~\cite{Unruh95,Steane99,Ladd10}. By capitalizing on the fundamental properties of photon-matter interaction in the solid state, the field of circuit quantum electrodynamics (cQED)~\cite{You1,Blais04,Wallraff04,You2,You3,Blais07,Schuster07,DiCarlo09,Clarke08,Buluta1,Buluta2,Niemczyk10,Xiang1} provides an attractive approach to removing this obstacle. Recent advances in the field have led to the realization of quantum bits, qubits, approaching the requirements set by quantum threshold theorems~\cite{Paik11,Devoret13} and shown that coupling a qubit to a coplanar-waveguide (CPW) resonator can offer protection against unwanted noise~\cite{Koch07,Manucharyan09,Nataf11}. A promising demonstration of the control over the system-bath interaction provided by the cQED architecture is the proposal for studying single-photon heat conduction~\cite{CavRevI,CavRevII}. Entering a regime where such a conduction mechanism is dominant requires accurate tuning of the system, accompanied by extreme robustness against noise.

A recent proposal introduced a cQED setup in which the coupling between a superconducting qubit and a thermal environment may be tuned at will, allowing for either rapid and precise initialization or typical protected evolution of the qubit~\cite{scirep3.1987}. Such a setup offers several advantages over previous works~\cite{Johnson12,Riste12,Reed10} and is, in principle, realizable with current technologies. Instead of exploiting a measurement feedback approach to control the coupling to the electromagnetic environment~\cite{Johnson12,Riste12} or modifying the decay channels~\cite{Reed10}, the setup realizes a complementary passive protocol for tuning the coupling to a single bosonic bath. The setup is based on a CPW transmission line divided into two capacitively coupled cavities, one housing a resistor and the other housing a qubit. By carefully selecting the system parameters, we can enter a regime where the resistor acts as the dominant noise source for the qubit and the coupling between the two can be externally tuned by manipulating the Josephson inductance of a set of superconducting quantum interference devices (SQUIDs) placed inside the cavity. The proof of principle for this tunable environment exploited the distributed element (DE) model, wherein the CPW cavity is represented by an infinite number of differential capacitors and inductors~\cite{Pozar04}. The model also allows one to account for the positioning of the resistor and the qubit inside the cavity. However, it would be beneficial to find a model system exhibiting similar tuning characteristics, but in which the complete dynamics are more easily accessible and some of the requirements of the DE model are lifted.

In this paper, we study a setup where the CPW cavities are replaced by a pair of coupled quantum LC oscillators. In our analysis, and unlike in Ref.~\cite{scirep3.1987}, the resistor is capacitively rather than galvanically coupled to one of the oscillators which is an advantageous property in the fabrication point of view of such devices. The result of this study is a simple and intuitive model for a qubit with an externally tunable coupling to a dissipative environment over many orders of magnitude. The model is designed as a proof of principle and accounts for a coupling to a single resistive bath omitting additional intrinsic noise sources. These intrinsic and possible external high-temperature noise sources are typically weakly coupled to the qubit, yet possibly raising its effective temperature substantially. By strongly coupling a low-temperature resistive bath to the qubit, it dominates over other noise sources lowering the effective qubit temperature significantly which allows for a more efficient initialization protocol. This adds to the toolbox of nanoscale temperature control~\cite{Muhonen12,Giazotto06,Valenzuela06,You4,Grajcar1,Grajcar2}. Furthermore, the model system has an analytical solution for several experimentally important cases. As a means of comparison, we establish a meaningful mapping between the model system studied here and the DE model system, and demonstrate that we can arrive at the same eigenspectrum. 

The outline of the paper is as follows. In Sec.~\ref{sec:model}, we introduce our physical system. Analytical solutions for the low-energy dynamics are given and analyzed in Sec.~\ref{sec:analytical} with special attention devoted to the protection of the qubit. In Sec.~\ref{sec:tuning}, we introduce the operating principle of our system and study the control over the qubit lifetime. Section~\ref{sec:mapping} presents a mapping between the CPW cavity model system and our model system. We compare the excitation spectra between the models and study the emerging effective resonator-qubit coupling capacitance. We conclude the paper in Sec.~\ref{sec:conclusions}.

\section{System} \label{sec:model}

\begin{figure}[]
\includegraphics[width=0.48\textwidth]{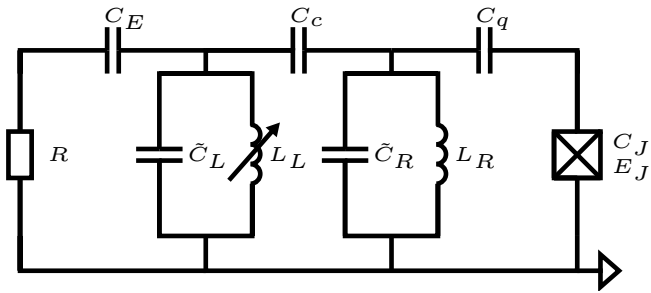}
\caption{Effective circuit diagram of our model system. The left and right LC resonators are coupled by a capacitor $C_c$. A resistor of resistance $R$ is capacitively coupled to the left resonator via a capacitance $C_{E}$ and acts as a noise source. The right resonator is coupled via capacitance $C_q$ to a Cooper pair box whose Josephson capacitance is $C_J$ and Josephson energy $E_J$.}
\label{fig:setup}
\end{figure}

Our system includes a sequence of lumped-element LC resonators. For simplicity, we restrict our analysis here to two such resonators and denote them as left and right corresponding to the schematic representation of the circuit given in Fig.~\ref{fig:setup}.
Each resonator is comprised of a lumped inductor of inductance $L_{k}$ and a lumped capacitor of capacitance $\tilde{C}_{k}$, where $k = L,R$ refers to the left and right resonators. The resonators are coupled through a capacitance $C_c$. To introduce a coupling to a bosonic environment, a resistor of resistance $R$ is capacitively coupled to the left resonator via the capacitance $C_{E}$. The right resonator is coupled to a Cooper pair box (CPB) operated as a qubit in the transmon regime~\cite{Koch07}. This coupling capacitance is denoted by $C_q$, and the Josephson junction defining the superconducting island has capacitance $C_J$ and Josephson energy $E_J$. In experimental realizations of the system, the inductance of the left resonator is made tunable by inserting one or more SQUIDs, whose Josephson inductances are tuned by external magnetic fields, in series with the inductor. Such insertion generally adds dissipation channels but assuming a strong coupling to the resistive environment, these additional channels can be neglected.

On the general level, the working principle of our system described in Fig.~\ref{fig:setup} is the following. The left resonator acts as a tunable filter for the noise and dissipation arising from the resistor. Thus if the left resonator is tuned in resonance with the qubit, the noise at this frequency is transmitted through the left resonator and introduces dissipation for the qubit. If the left resonator is in resonance with the right resonator, strong dissipation is introduced there. If the left resonator is far off resonance with either the right resonator or the qubit, the noise is heavily filtered and the effect of the resistor on the dynamics of the corresponding element is very weak.

The Hamiltonian for our model can be written as the sum of the charging, inductive, resistive, and Josephson Hamiltonians corresponding to each element in Fig.~\ref{fig:setup}. If we neglect terms of second order in the resistor voltage fluctuations, the Hamiltonian becomes
\begin{equation}
\hat{H}_{\mathrm{tot}} = \hat{H}_L + \hat{H}_R + \hat{H}_{L-R} + \hat{H}_{\mathrm{int}} + \hat{H}_{\mathrm{res}} + \hat{H}_q + \hat{H}_{R-q},
\label{eq:H2}
\end{equation}
where each term describes a subsystem or a coupling between them. The resonator Hamiltonians obtain an effective form due to the internal system couplings
\begin{equation}
\hat{H}_k = \frac{1}{2} C_k \hat{V}_k^2 + \frac{1}{2} L_k \hat{I}_k^2, \qquad k=L,R,
\label{eq:Hk}
\end{equation}
where $\hat{V}_k$ is the operator for the voltage across the $k$th LC circuit, with $\hat{I}_k$ the corresponding current operator. The effective capacitances are $C_L = \tilde{C}_L + C_c+C_E$ and $C_R = \tilde{C}_R + C_c + C_q$. The presentation in Eq.~(\ref{eq:Hk}) is chosen due to its practical descriptiveness and can be transformed to use the typical canonically conjugate operators for charge and magnetic flux~\footnote{When $C_E, C_c, C_q \ll \tilde{C}_L, \tilde{C}_R, C_J$, Eq.~(\ref{eq:Hk}) can be rewritten using the operators for charge $\hat{Q}_k = C_k \hat{V}_k$ and magnetic flux $\hat{\Phi}_k = L_k \hat{I}_k$ of the $k$th LC resonator. These operators are canonically conjugate fulfilling the commutation relation $[\hat{Q}_k,\hat{\Phi}_k] = i\hbar$.}. The resonator-resonator coupling term is
\begin{equation}
\hat{H}_{L-R} = - C_c\hat{V}_L\hat{V}_R,
\end{equation}
and, similarly, the coupling between the resonators and the resistor assumes the form
\begin{equation}
\hat{H}_{\mathrm{int}} = - C_E \hat{V}_L \delta \hat{V}_{\mathrm{res}},
\label{eq:Hint}
\end{equation}
where $\delta \hat{V}_{\mathrm{res}}$ describes the intrinsic voltage fluctuations across the resistor~\footnote{The resonator-resistor coupling Hamiltonian, $\hat{H}_{\mathrm{int}}$, is determined by the operator for the voltage over $C_E$ and, hence, depends on the intrinsic voltage fluctuations $\delta \hat{\tilde{V}}_{\mathrm{res}}$ over the resistor circuit. Accounting for the total impedance of the circuit, the spectral density of the voltage fluctuations across $C_E$ is given by $S_{\delta \tilde{V}_{\mathrm{res}}}(\omega) = [1/(\omega C_E)]^2/|R+1/(i\omega C_E)|^2 S_{\delta V_{\mathrm{res}}}(\omega)$, where $S_{\delta V_{\mathrm{res}}}(\omega)$ is the spectral density of the voltage fluctuations $\delta \hat{V}_{\mathrm{res}}$ over the resistor. In this paper, we select the system parameters such that $\omega \ll 1/(RC_E)$ implying that $S_{\delta \tilde{V}_{\mathrm{res}}}(\omega) \approx S_{\delta V_{\mathrm{res}}}(\omega)$, and the Hamiltonian in Eq.~(\ref{eq:Hint}) accurately approximates the resonator-resistor coupling. This is essentially the same approximation that is employed in writing down Eq. (\ref{eq:H2}) by neglecting the terms that are of the second order in the voltage fluctuations.}. The resistor Hamiltonian, $\hat{H}_{\mathrm{res}}$, represents a bosonic thermal bath~\cite{Caldeira81}. The resistor can also be modelled as a transmission line~\cite{Yurke} although this approach is not employed here. The qubit Hamiltonian consists of the charging and Josephson parts~\cite{Vion02}
\begin{equation}
\hat{H}_q = \frac{1}{2} (C_J + C_q) \hat{V}_q^2 + E_J \cos \hat{\phi}, 
\label{eq:Hq}
\end{equation}
where $\hat{V}_q$ and $\hat{\phi}$ are the operators for the voltage and superconducting phase difference across the junction, and for simplicity, we have neglected the self-capacitance of the superconducting island. Similarly to the resonator Hamiltonians, Eq.~(\ref{eq:Hq}) can be given in a more typical presentation using the Cooper-pair number operator for the junction instead of the operator for the voltage across it~\footnote{When $C_E, C_c, C_q \ll \tilde{C}_L, \tilde{C}_R, C_J$, we can define $\hat{N} = -(C_J + C_q)\hat{V}_q/(2e)$ as the Cooper-pair number operator for the junction transferring to a more typical presentation. The phase difference operator $\hat{\phi}$ is canonically conjugate to $\hat{N}$ fulfilling $[\hat{\phi},\hat{N}] = i$.}. The coupling between the right resonator and the qubit is given by
\begin{equation}
\hat{H}_{R-q} = -C_q \hat{V}_R \hat{V}_q.
\end{equation}

The resonator Hamiltonians given in Eq.~(\ref{eq:Hk}) are of the form of quantum harmonic oscillators and are diagonalized by transforming to use the canonically conjugate operators for charge and magnetic flux followed by an introduction of $\hat{a}_k^{\dagger}$ ($\hat{a}_k$) as the relevant bosonic creation (annihilation) operator for the $k$th resonator. In our case, the coupling capacitances are assumed to be much smaller than other system capacitances, that is, $C_E, C_c, C_q \ll \tilde{C}_L, \tilde{C}_R, C_J$, and hence the diagonalization yields 
\begin{equation}
\hat{H}_k = \hbar \omega_k(\hat{a}_k^{\dag}\hat{a}_k+1/2 ),
\end{equation}
where $\omega_k=1/\sqrt{L_k C_k}$, and allows us to define the voltage operators as $\hat{V}_{k}=V^{0}_{k}(\hat{a}^{\dag}_{k} + \hat{a}_{k})$, where $V^{0}_{k}=\sqrt{\hbar \omega_{k}/(2 C_{k})}$, and the current operators as $\hat{I}_k = I_k^0 (\hat{a}_k^{\dagger}-\hat{a}_k)$, where $I_k^0 = i \sqrt{\hbar \omega_k /(2L_k)}$. Using the same formalism, the resonator-resonator coupling term can be written as 
\begin{equation}
\hat{H}_{L-R}=\hbar \alpha(\hat{a}^{\dag}_{L}\hat{a}^{\dag}_{R}+\hat{a}^{\dag}_{L}\hat{a}_{R}+\hat{a}_{L}\hat{a}^{\dag}_{R}+\hat{a}_{L}\hat{a}_{R}),
\end{equation}
where we define $\alpha=-C_c V^{0}_{L}V^{0}_{R}/\hbar$ as the resonator-resonator coupling strength. Assuming that the qubit is operated in the transmon regime, the operator for the voltage across the junction is $\hat{V}_q=-i\frac{\sqrt{2}e}{C_J+C_q}\left(\frac{E_J}{8E_C}\right)^{1/4}(\hat{b}-\hat{b}^{\dag})$, with $E_C = e^2/[2(C_J+C_q)]$ being the charging energy for the CPB island and $\hat{b}^{\dag}$ and $\hat{b}$ being the creation and annihilation operators for the harmonic oscillator approximating the transmon, respectively~\cite{Koch07}. Hence, the resonator-qubit coupling Hamiltonian becomes
\begin{equation} 
\hat{H}_{R-q} = \hbar g(\hat{b}\hat{a}^{\dag}_{R}+\hat{b}\hat{a}_{R}-\hat{b}^{\dag}\hat{a}_{R}-\hat{b}^{\dag}\hat{a}^{\dag}_{R}),
\label{eq:Hresqb}
\end{equation}
where $g=i\frac{\sqrt{2}eC_qV_R^0}{\hbar (C_J+C_q)}\left(\frac{E_J}{8E_C}\right)^{1/4}$ is the resonator-qubit coupling strength.

Our main interest is the dynamics of the resonator-qubit system and, to this end, we treat the resistor as a dissipative environment described by the thermal bath Hamiltonian $\hat{H}_{\mathrm{res}}$. Assuming that the coupling to the environment is weak, the coupling term $\hat{H}_{\mathrm{int}}$ can be taken as a small perturbation inducing transitions between the resonator-qubit states defined by the eigenproblem $(\hat{H}_L + \hat{H}_R + \hat{H}_{L-R} + \hat{H}_q + \hat{H}_{R-q})\ket{m} = E_m \ket{m}$. The transition rates are approximated using the Fermi golden rule~\cite{Clerk10} as 
\begin{equation}
\Gamma_{m\rightarrow n} = \frac{|\bra{n} C_E \hat{V}_L \ket{m}|^{2}}{\hbar^{2}} S_{\delta V_{\mathrm{res}}}(-\omega_{mn}).
\label{eq:rate}
\end{equation}
Here, $\omega_{mn} = (E_n-E_m)/\hbar$ and we assume that the resistor is in thermal equilibrium so that the voltage fluctuations $\delta\hat{V}_{\mathrm{res}}$ follow the Johnson--Nyquist spectral density $S_{\delta V_{\mathrm{res}}}(\omega)=2 \hbar R \omega/[1-e^{-\hbar \omega/(k_{B}T)}]$, where $T$ is the effective resistor temperature. In the physical implementation of the device presented in Fig.~\ref{fig:setup}, additional intrinsic noise sources may emerge that are not accounted for by our model. However, we assume that their effect can be greatly mitigated with current fabrication techniques so that the main features of tunability provided by our scheme can be achieved.

\section{Analytical results} \label{sec:analytical}

Attaining closed-form solutions for the eigenproblem presented in the previous section is likely not plausible in the general case and we therefore focus our efforts on a few important special cases in the low-energy regime. This analysis provides convenient tools for experimentalists working on the system and offers insight on the robustness of the qubit against noise.

\subsection{Decoupled qubit} \label{subsec:ics}
 
We begin by studying the dynamics of the system shown in Fig.~\ref{fig:setup} when the qubit has been completely decoupled, i.e., when $C_q = 0$. We define the excitation number product state basis as $\{\ket{n_L,n_R}\}$, where $n_L$ and $n_R$ are the excitation numbers of the left and right resonators, respectively. Furthermore, we consider only the low-energy dynamics so that the basis is restricted to $\{ \ket{0,0},\ket{0,1},\ket{1,0} \}$, omitting simultaneous excitations of the left and right resonators. We choose the zero-point energy so that the lowest eigenenergy is $E_0=0$ corresponding to the ground state $\ket{g}=\ket{0,0}$. The subsequent excited state energies are $E_{\pm}= \hbar( \Omega \pm \sqrt{\Delta^{2}+\alpha^{2}})$, where $\Omega = (\omega_L+\omega_R)/2$ and $\Delta = (\omega_L- \omega_R)/2$, and the excited states are
\begin{align}
&\ket{-}=-\cos(\theta)\ket{1,0}+\sin(\theta) \ket{0,1}, \label{eq:decqubitstates1} \\
&\ket{+}=\sin(\theta)\ket{1,0}+\cos(\theta) \ket{0,1}, \label{eq:decqubitstates2}
\end{align}
where $\tan(\theta)=(\Delta + \sqrt{\Delta^{2}+\alpha^{2}})/\alpha$. With the help of Eqs.~(\ref{eq:rate})--(\ref{eq:decqubitstates2}), the relaxation rates can be written as
\begin{align}
\Gamma_{- \rightarrow g} &= \frac{\alpha^{2}}{[(\Delta+\sqrt{\Delta^{2}+\alpha^{2}})^{2}+\alpha^{2}]} \frac{R \omega_{-}}{Z_{L}[1-e^{-\hbar \omega_{-}/(k_{B}T)}]}\left(\frac{C_E}{C_L} \right)^{2}, \\
\Gamma_{+ \rightarrow g} &= \frac{(\Delta+\sqrt{\Delta^{2}+\alpha^{2}})^{2}}{[(\Delta+\sqrt{\Delta^{2}+\alpha^{2}})^{2}+\alpha^{2}]} \frac{R \omega_{+}}{Z_{L}[1-e^{-\hbar \omega_{+}/(k_{B}T)}]}\left(\frac{C_E}{C_L} \right)^{2},
\end{align}
and the corresponding excitation rates are acquired through a substitution $\Gamma_{g \rightarrow \pm}(\omega)=\Gamma_{\pm \rightarrow g}(-\omega)$. Above, we defined the characteristic impedance of the $k$th resonator as $Z_k=\sqrt{L_{k}/C_{k}}$ and used a notation $\omega_{\pm} = E_{\pm}/\hbar$. Near resonance in which $\Delta \rightarrow 0$, we have $\sin(\theta) \rightarrow 1/\sqrt{2}$ and $\cos(\theta) \rightarrow 1/\sqrt{2}$, and hence $\Gamma_{+ \rightarrow g} \approx \Gamma_{- \rightarrow g}$ for $\alpha \ll \Omega$. The rates at the resonance are $\Gamma_{\pm \rightarrow g}^{\mathrm{res}} = R (\Omega \pm \alpha) (C_E/C_L)^2 / [2Z_L (1-e^{-\hbar (\Omega \pm \alpha)/(k_{B}T)})]$ and the eigenstates become equal superpositions of the excited resonator states. 

In the far-detuned limit $\Delta \gg \alpha$, we have $\cos^2(\theta) \approx \alpha^{2}/(4\Delta^{2}) $ and $\sin^2(\theta) \approx 1-\alpha^{2}/(4\Delta^{2})$, in the case of which the transition rates reduce to
\begin{align}
\Gamma_{- \rightarrow g}^{\mathrm{far}} &\approx \frac{\alpha^{2}}{4\Delta^{2}}\frac{R \omega_{-}}{ Z_{L}[1-e^{-\hbar \omega_{-}/(k_{B}T)}]} \left(\frac{C_E}{C_L} \right)^{2}, \\
\Gamma_{+ \rightarrow g}^{\mathrm{far}} &\approx \left( 1-\frac{\alpha^{2}}{4\Delta^{2}}\right)\frac{R \omega_{+}}{ Z_L[1-e^{-\hbar \omega_{+}/(k_{B}T)}]} \left(\frac{C_E}{C_L} \right)^{2}.
\end{align}
The eigenstate $\ket{-}$ approximately corresponds to an excitation in the right resonator whereas the eigenstate $\ket{+}$ corresponds to an excitation in the left resonator. To quantify the effect of the detuning, we can define the ratio between the decay rate of an excited state in Eqs.~(\ref{eq:decqubitstates1})--(\ref{eq:decqubitstates2}) far from resonance and at resonance, the tuning ratio $r_{\pm} = \Gamma_{\pm \rightarrow g}^{\mathrm{far}} / \Gamma_{\pm \rightarrow g}^{\mathrm{res}}$. For the state $\ket{-}$ approximately corresponding to the excited state of the right resonator far from resonance, the ratio is approximately $r_- \approx \alpha^2/(2\Delta^2)$ for $\alpha, \Delta \ll \Omega$. Hence we can effectively decouple the right resonator from the noise source by detuning. As a consequence, a qubit directly coupled only to the right resonator can be efficiently protected.

\subsection{Resonator-qubit system} \label{subsec:cqs}

We proceed by analyzing a few important regimes where a closed-form solution for the full system in Fig.~\ref{fig:setup} is accessible. This analysis not only sets the premise for studying the general behavior of the system, but also offers simple tools for experimentalists operating in these important regimes. We extend the computational basis from the previous section to include the qubit so that the basis states are of the form $\ket{n_L, n_R, \sigma}$, where $\sigma = g,e$ for the ground and excited qubit states, respectively. The basis relevant to the low-energy dynamics becomes $\{ \ket{0,0,g},\ket{1,0,g},\ket{0,1,g},\ket{0,0,e} \}$ which does not include states of the form of tensor products of the excited qubit state and resonator excitations as those only couple to higher-order photonic excitations. This omission invokes effectively the rotating wave approximation so that the resonator-qubit interaction is of the form $\hat{H}_{R-q} = \hbar (g \hat{a}_R^{\dagger} \hat{\sigma}^- + g^* \hat{a}_R \hat{\sigma}^+)$, where $\hat{\sigma}^+$ and $\hat{\sigma}^-$ are the Pauli raising and lowering operators for the qubit, respectively. Within the selected basis, the time-independent Schr\"odinger equation for the resonator-qubit system can be written in a practical form yielding both the eigenenergies and eigenvectors [see Appendix~\ref{app:analyticalsolution}]. 

First, we study the partial resonance case where $\omega_R \ne \omega_L = \omega_q = \omega$, that is, the qubit is in resonance with the dissipative resonator but not with the one to which it is directly coupled. Here $\hbar \omega_q$ denotes the qubit excitation energy obtained by diagonalizing $\hat{H}_q$. Using Appendix~\ref{app:analyticalsolution}, the excited state energies $\{ \epsilon_k \}$ obtain closed-form solutions as $\epsilon_1 = \hbar \omega$, $\epsilon_{2/3} = \hbar( \omega_\textrm{av} \pm \sqrt{\omega_\textrm{d}^{2}+|g|^{2}+\alpha^{2}})$, where $\omega_\textrm{av}=(\omega+\omega_R)/2$ and $\omega_\textrm{d}=(\omega-\omega_R)/2$. The corresponding eigenvectors are
\begin{align}
&\ket{\varphi_1} = \frac{\alpha}{\sqrt{\alpha^{2}+|g|^{2}}} \left(\frac{g}{\alpha}\ket{1,0,g}+\ket{0,0,e}\right), \label{eq:anresstate1}  \\
&\ket{\varphi_2} = \frac{|g|}{\Omega_+}\left\{\frac{\alpha}{g}\ket{1,0,g}+i\left[A+\left(\frac{\omega_d}{|g|}\right)\right]\ket{0,1,g}+\ket{0,0,e} \right\}, \\
&\ket{\varphi_3} = \frac{|g|}{\Omega_-}\left\{\frac{\alpha}{g}\ket{1,0,g}-i\left[A- \left(\frac{\omega_d}{|g|}\right)\right]\ket{0,1,g}+\ket{0,0,e} \right\}, \label{eq:anresstate3}
\end{align}
where $A=\sqrt{\left(\omega_d/|g|\right)^{2}+(\alpha/|g|)^2+1}$ and $\Omega_{\pm}=\sqrt{\alpha^{2}+|g|^{2}+(A|g| \pm \omega_d)^{2}}$. Note that the ground state is $\ket{\varphi_0} = \ket{0,0,g}$ with the zero-point in energy chosen so that $\epsilon_0 = 0$. Using Eqs.~(\ref{eq:anresstate1})-(\ref{eq:anresstate3}), the relaxation rates to the ground state become
\begin{align}
&\Gamma_{\varphi_1 \rightarrow \varphi_0}^{\mathrm{qres}} =  \frac{|g|^{2}}{|g|^{2}+\alpha^{2}}\frac{R \omega_{1}}{ Z_{L}[1-e^{-\hbar \omega_{1}/(k_{B}T)}]} \left(\frac{C_E}{C_L} \right)^{2}, \\
&\Gamma_{\varphi_2 \rightarrow \varphi_0}^{\mathrm{qres}} =  \frac{\alpha^{2}}{\Omega_{-}^2}\frac{R \omega_{2}}{ Z_L[1-e^{-\hbar \omega_{2}/(k_{B}T)}]} \left(\frac{C_E}{C_L} \right)^{2}, \\
& \Gamma_{\varphi_3 \rightarrow \varphi_0}^{\mathrm{qres}} = \frac{\alpha^{2}}{\Omega_{+}^2}\frac{R \omega_{3}}{ Z_L[1-e^{-\hbar \omega_{3}/(k_{B}T)}]} \left(\frac{C_E}{C_L} \right)^{2}, \label{eq:resrate3}
\end{align}
where $\omega_{m} = \epsilon_m/\hbar$, $m=1,2,3$, for simplicity. The excitation rates are obtained from $\Gamma_{\varphi_0 \rightarrow \varphi_m}^{\mathrm{qres}}(\omega)=\Gamma_{\varphi_m \rightarrow \varphi_0}^{\mathrm{qres}}(-\omega)$. The full resonance solution is obtained by setting $\omega_R = \omega$ and if we additionally set $\alpha=|g|$, the rates satisfy $\Gamma_{\varphi_1 \rightarrow \varphi_0}^{\mathrm{res}} \approx 2\Gamma_{\varphi_2 \rightarrow \varphi_0}^{\mathrm{res}} \approx 2\Gamma_{\varphi_3 \rightarrow \varphi_0}^{\mathrm{res}}$, when $\sqrt{2}\alpha \ll \omega$, and obtain the form $\Gamma_{\varphi_1 \rightarrow \varphi_0}^{\mathrm{res}} = R \omega (C_E/C_L)^2 / [2Z_{L}(1-e^{-\hbar \omega/(k_{B}T)})]$ and  $\Gamma_{\varphi_k \rightarrow \varphi_0}^{\mathrm{res}} = R (\omega \pm \sqrt{2}\alpha) (C_E/C_L)^2 / [4Z_{L}(1-e^{-\hbar (\omega\pm\sqrt{2}\alpha)/(k_{B}T)})]$ with plus and minus corresponding to $k=2$ and $k=3$, respectively.

If we study the case where the resonator-resonator coupling is much stronger than the resonator-qubit coupling, i.e., $\alpha \gg |g|$, the closed-form solutions indicate that the qubit decay is much slower than the decay of the resonators. Here we identified $\ket{\varphi_1}$ as the excited state of the qubit and $\ket{\varphi_2}$ and $\ket{\varphi_3}$ as superpositions of the excited states of the resonators by studying the corresponding state amplitudes. If $\alpha \ll |g|$, on the other hand, the decay of the left resonator can be seen dominate over both the qubit decay and the decay of the right resonator. Similarly to the previous case, we identified $\ket{\varphi_1}$ as the excited state of the left resonator and $\ket{\varphi_2}$ and $\ket{\varphi_3}$ as superpositions of the excited states of the right resonator and the qubit. Note that these observations in the resonance case serve as a consistency check for the system and are not to be confused with the control introduced by manipulating the resonator-resonator and resonator-qubit detunings. If $\alpha$ and $|g|$ are of the same order of magnitude, the detunings determine the relative decay rates as we will show in the following section.

Next, we study the case where the system couplings are weak $|\alpha|,|g| \ll \omega_L,\omega_R,\omega_q$ and the system is far away from resonance $|\alpha|, |g| \ll |\omega_{l}-\omega_{s}|$, for all $l,s \in \{L,R,q\}$, so that a perturbation theory solution becomes available. After expanding to second order in the small coupling parameters, the resulting eigenenergies and states are given in Appendix~\ref{app:weak}. Using these results, the relaxation to the ground state takes place at rates defined by
\begin{align}
&\Gamma_{\varphi_1 \rightarrow \varphi_0}^{\mathrm{far}} =  A_1^{2} \left(\frac{2\Delta_{LR}^{2}-\alpha^{2}}{2\Delta_{LR}^{2}}\right)^{2}\frac{R \omega_{1}}{ Z_{L}[1-e^{-\hbar \omega_{1}/(k_{B}T)}]} \left(\frac{C_E}{C_L} \right)^{2}, \label{eq:weakrate1} \\
&\Gamma_{\varphi_2 \rightarrow \varphi_0}^{\mathrm{far}} =  A_2^{2} \frac{\alpha^{2}}{\Delta_{RL}^{2}}\frac{R \omega_{2}}{ Z_L[1-e^{-\hbar \omega_{2}/(k_{B}T)}]} \left(\frac{C_E}{C_L} \right)^{2}, \label{eq:weakrate2} \\
& \Gamma_{\varphi_3 \rightarrow \varphi_0}^{\mathrm{far}} = A_3^2 \frac{|g|^2 \alpha^2}{\Delta_{qR}^2\Delta_{qL}^2}\frac{R \omega_{3}}{ Z_L[1-e^{-\hbar \omega_{3}/(k_{B}T)}]} \left(\frac{C_E}{C_L} \right)^{2}. \label{eq:weakrate3}
\end{align}
Here we denote $\Delta_{ls} = \omega_l-\omega_s$ as the difference between two intrinsic system frequencies and the normalization constants $A_k$ are given in Eqs.~(\ref{eq:weakN1})--(\ref{eq:weakN3}). Note that $A_k = 1 + O(|\alpha|/|\Delta_{ls}|,|g|/|\Delta_{ls}|)$, for all $k=1,2,3$ and $l,s = L,R,q$. Assuming $|\alpha|, |g| \ll |\omega_{l}-\omega_{s}|$ implies that $A_k \approx 1$ making Eqs.~(\ref{eq:weakrate1})--(\ref{eq:weakrate3}) easier to interpret. We observe that the relaxation rate of the state $\ket{\varphi_3}$ ($\approx \ket{0,0,e})$ is proportional to $|g|^2 \alpha^2$, that is, the excited qubit state is \emph{doubly protected} against noise: firstly by the weak resonator-qubit coupling and secondly by the weak resonator-resonator coupling. This is in contrast to state $\ket{\varphi_2}$ ($\approx \ket{0,1,g}$) which is proportional to $\alpha^2$ and, hence, only benefits from the weak resonator-resonator coupling. Similarly to Sec.~\ref{subsec:ics}, we can define the tuning ratio $r_{\varphi_3} = \Gamma_{\varphi_3 \rightarrow \varphi_0}^{\mathrm{far}} / \Gamma_{\varphi_3 \rightarrow \varphi_0}^{\mathrm{qres}}$ for the state $\ket{\varphi_3}$ approximating the excited state of the qubit far away from partial resonance. Using Eq.~(\ref{eq:resrate3}) for the rate at partial resonance and Eq.~(\ref{eq:weakrate3}) for the rate far away from resonance, we obtain $r_{\varphi_3} \approx |g|^2/\Delta_{qL}^2$ for $|\omega_{\mathrm{d}}| \ll |\omega_{\mathrm{av}}|$ and weak system couplings.

\section{Tuning the qubit and photon lifetimes} \label{sec:tuning}

The control over the qubit lifetime stems from the resonators intermediating the relevant interaction with the bath. This becomes clear when we assume that the inductance of the left resonator can be externally controlled in the physical realization of the system. For instance, if the resonator describes a CPW cavity as discussed in Sec.~\ref{sec:mapping}, such manipulation can be implemented by the insertion of SQUIDs into the cavity. Manipulating the frequency of the left resonator in this manner effectively allows us to drag the left resonator on and off resonance, $\omega_L = \omega_q$ being the resonance condition, with the qubit. A simple qualitative explanation for the resulting changes in the qubit lifetime can be found by considering the impedance of the LC resonators in a classical circuit~\cite{Horowitz:1989:AE:76734}. Far above resonance, the impedances of the resonators to the ground may be considered to be purely capacitive. Consequently the coupling-capacitor--resonator circuit acts as a voltage divider, resulting in effective voltage fluctuations which are reduced at the position of the qubit protecting it from decay. On resonance, the impedance of the resonator to the ground may be considered infinite so that the qubit-bath coupling is of the form of an effective series capacitance resulting in fast decay.

\begin{figure}
\includegraphics[width=0.5\textwidth]{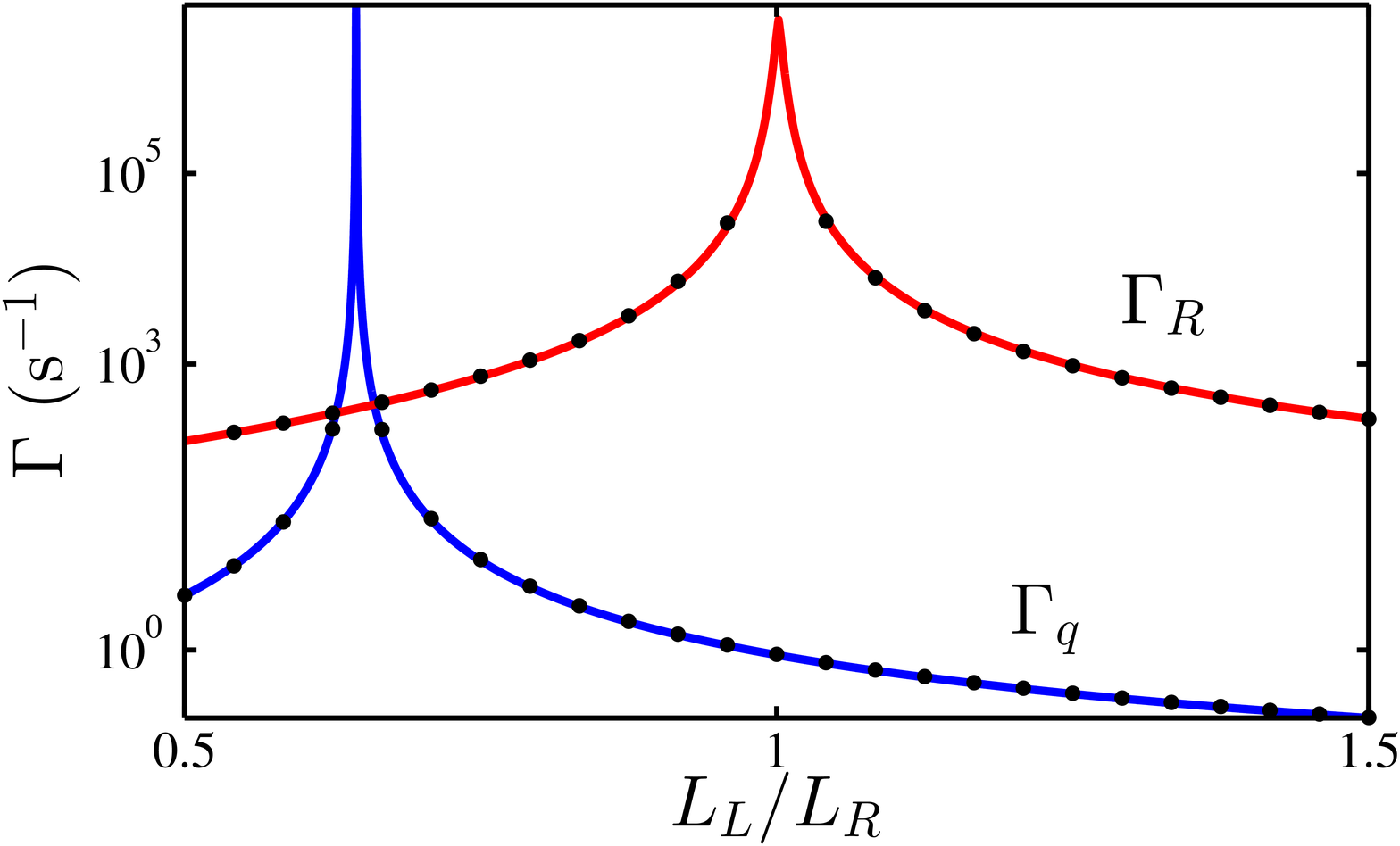}
\caption{(Color online) Relaxation rate to the ground state of the eigenstate $\ket{q}$, $\Gamma_q$, and the eigenstate $\ket{R}$, $\Gamma_R$, as a function of the inductance of the left resonator presented by the solid lines. Far from resonance, these rates are accurately approximated by the weak-coupling solutions of Eqs.~(\ref{eq:weakrate2})--(\ref{eq:weakrate3}) presented by the dotted lines. For the coupling capacitances, we use $C_c=C_q=C_E=1 \textrm{ fF}$. The resonators have capacitors of magnitude $\tilde{C}_L=\tilde{C}_R=39 \textrm{ pF}$, and the inductance of the right resonator is fixed to $L_R=40\textrm{ nH}$ resulting in $\omega_R/(2\pi)= 12.79\textrm{ GHz}$. The resistor has a resistance of $R = 500\textrm{ }\Omega$ and an effective temperature of $T = 10$ mK, and the qubit parameters are chosen such that $E_J/E_C=50$, while its frequency is fixed at $\omega_q/(2\pi)= 15.92\textrm{ GHz}$. The resonator-qubit coupling strength is $|g|/(2\pi) = 178.25 \textrm{ MHz}$ and the resonator-resonator coupling strength ranges from $\alpha/(2\pi) = -19.40 \textrm{ MHz}$ at $L_L/L_R = 0.5$ to $\alpha/(2\pi) = -14.74 \textrm{ MHz}$ at $L_L/L_R = 1.5$.}
\label{fig:rates}
\end{figure}

In the following simulation, we identify two of the low-energy eigenstates of the qubit-resonator system as the ones approximating $\ket{q} \approx \ket{0,0,e}$ far from the resonator-qubit resonance $\omega_q = \omega_L$, and $\ket{R} \approx \ket{0,1,g}$ far from the resonator-resonator resonance $\omega_R = \omega_L$. Even though these eigenstates do not match the excited states of the qubit and the right resonator near the resonances, they still allow us to study the operating principle of the system. We calculate the decay rate of $\ket{q}$ to the ground state, $\Gamma_q$, from Eq.~(\ref{eq:rate}) by numerically diagonalizing the Hamiltonian in Eq.~(\ref{eq:H2}), and present the rate in Fig.~\ref{fig:rates} as a function of the inductance of the left resonator.
The rate obtains a sharp peak when the left resonator is tuned near resonance with the qubit, as expected. Note that with the realistic system parameters used in Fig.~\ref{fig:rates}, tuning the inductance allows for significant control over the qubit lifetime. For instance, adjusting the resonance frequency of the left resonator from the point $\omega_L = \omega^{\mathrm{peak}}_q$, at which $\Gamma_q$ obtains its maximum value, to $\omega_L/\omega^{\mathrm{peak}}_q = 0.75$ yields a decrease in the decay rate given by the ratio $\Gamma_q (\omega^{\mathrm{peak}}_q) / \Gamma_q (0.75\omega^{\mathrm{peak}}_q) = 1.69 \times 10^6$. Hence, externally manipulating the inductance makes it possible to both rapidly initialize the qubit to the ground state and to protect it from dissipation. The frequency $\omega^{\mathrm{peak}}_q$ differs slightly from the qubit frequency $\omega_q$ due to the resonator-resonator coupling strength having a nonlinear dependence on $\omega_L$. The analytical solutions for when the system couplings are weak given in Eqs.~(\ref{eq:weakrate1})--(\ref{eq:weakrate3}) accurately approximate the numerical solutions far away from resonance in Fig.~\ref{fig:rates}. In particular, the rate $\Gamma_q$ is accurately approximated by $\Gamma_{\varphi_3 \rightarrow \varphi_0}^{\mathrm{far}}$ in Eq.~(\ref{eq:weakrate3}) near the resonator-resonator resonance despite the formal assumption that the system remains far away from all resonances used when deriving the analytical solutions. Even though nonidealities in the experimental realization of the system as well as other dissipation mechanisms can limit the physical rates in practice, the decay caused by the coupling to the thermal bath can be efficiently controlled in this manner.

We show also the decay rate of $\ket{R}$ to the ground state, $\Gamma_R$, in Fig.~\ref{fig:rates}. Similarly to the qubit case, this decay rate exhibits a strong peak which, however, resides at $\omega_L=\omega_R^\textrm{peak}$ near the resonator-resonator resonance $\omega_L = \omega_R$. Comparing the decay rates using the same relative changes in the frequency of the left resonator ($\omega_L/\omega^{\mathrm{peak}}_R: 1 \rightarrow 0.75$) yields a ratio of $\Gamma_R (\omega^{\mathrm{peak}}_R) / \Gamma_R (0.75\omega^{\mathrm{peak}}_R) = 2.66 \times 10^4$ where $\omega_L = \omega^{\mathrm{peak}}_R$ is the frequency at which $\Gamma_R$ reaches its maximum value. Even though the range of control is more limited than in the qubit case, we can still adjust the decay rate over several orders of magnitude. The difference in the ranges is an indication of the double protection of the qubit against noise mentioned in Sec.~\ref{subsec:cqs}. According to Eqs.~(\ref{eq:weakrate2})--(\ref{eq:weakrate3}), the change in both $\Gamma_q$ and $\Gamma_R$ is approximately proportional to $\alpha^2$ as the inductance of the left resonator is manipulated far from resonance. However, $\Gamma_q$ has an additional constant prefactor $|g|^2$ allowing for stronger protection and, hence, a larger range. By setting $\omega_q=\omega_L=\omega_R$, we can rapidly initialise the entire system. Alternatively, by detuning both the resonators, as well as the qubit, we can protect both the excited right resonator state, and the qubit from decay.

\section{Mapping from the distributed element model} \label{sec:mapping}

\begin{figure}[h]
\includegraphics[width=0.42\textwidth]{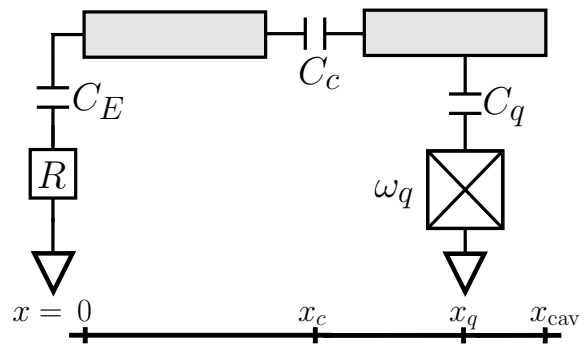}
\caption{Schematic representation of the CPW cavity system. The center conductor of the cavity divided by the coupling capacitor $C_c$ placed at $x=x_c$ is represented by the shaded elements. For clarity, the ground plane surrounding the center conductor is not shown. The capacitive bath coupling is introduced by coupling the left region via $C_E$ to the resistor $R$. The right cavity region is coupled via $C_q$ to a CPB defining a qubit with bare excitation frequency $\omega_q$. The qubit location $x=x_q$ represents the position at which the qubit is connected between the center conductor and the ground plane of the cavity.}
\label{fig:cavity}
\end{figure}

In Ref.~\onlinecite{scirep3.1987}, a similar-purpose physical system is considered as here but composing of a coplanar-waveguide cavity and a qubit. The cavity is separated into two distinct regions by inserting a capacitor of capacitance $C_c$ into the central conducting strip. These regions then house the resistor on the left and the qubit on the right along with an array of SQUIDs used to manipulate the inductance of the left region. The application of the distributed element (DE) model to this situation allows one to account for the positioning of the resistor and the qubit anywhere in the divided cavity. The key difference between this system and the one presented here is that the resistor is galvanically, as opposed to capacitively, coupled to the transmission line and therefore couples through the current rather than the voltage of the cavity. However, the setup using the DE model can be modified to incorporate a capacitive bath coupling physically justifying the search for a meaningful mapping between the models. We present a simplified schematic of the cavity system with a capacitive bath coupling in Fig.~\ref{fig:cavity}.

\begin{figure}[h]
\includegraphics[width=0.5\textwidth]{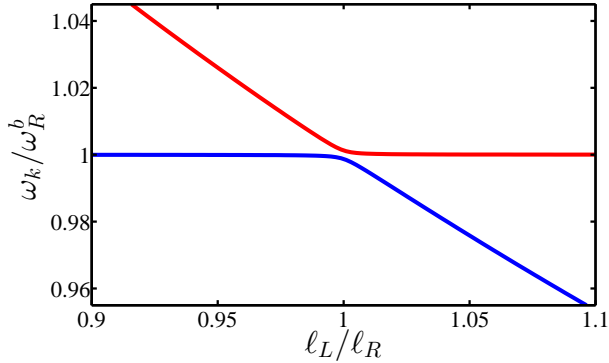}
\caption{(Color online) Excited state eigenfrequencies $\omega_k = \epsilon_k/\hbar$, $k=1,2$ (from bottom to top), as a function of the left cavity inductance per unit length. The eigenenergies $\epsilon_k$ are calculated for the cavity system ignoring both the resistor and the qubit, and they are equal to those calculated for our system using the mapping detailed in this section. The eigenfrequencies are scaled by the frequency of the bare right cavity $\omega_R^b = \pi/[(x_\mathrm{cav}-x_c)\sqrt{\ell_R c}]$. We use $x_\textrm{cav}=12\textrm{ mm}$, $x_c = x_\textrm{cav}/2$ and $C_c = 1\textrm{ fF}$. The cavity has a capacitance per unit length of $c=130\textrm{ pF/m}$, and the right cavity region has a characteristic impedance of $Z_c = \sqrt{\ell_R/c} = 50\textrm{ }\Omega$ and a bare frequency of $\omega^b_R/(2\pi)  = 12.82 \textrm{ GHz} $.}
\label{fig:spectrum}
\end{figure}

The task at hand is to select the parameters such that the two systems accurately exhibit similar physical properties. In the following, we refer to the resonators describing the individual cavity regions as bare cavities. In order for the DE and LE models to yield the same spectrum for the isolated resonator and cavity systems, we match the frequencies of the lowest bare cavity modes to the resonator frequencies in our model, and the corresponding voltage operators for the bare modes at the ends of the cavity regions to the resonator voltage operators in our model [see Appendix~\ref{app:baremapping}]. This yields a mapping given by
\begin{align}
\tilde{C}_L &= \frac{c x_c}{2}, \label{eq:bare1} \\ 
L_L& = \frac{2\ell_L x_c}{\pi^{2}}, \\  
\tilde{C}_R &= \frac{c (x_\textrm{cav}-x_c)}{2}, \\ 
L_R&=\frac{2\ell_R (x_\textrm{cav}-x_c)}{\pi^{2}}, \label{eq:bare4}
\end{align} 
where $c$ is the constant cavity capacitance per unit length, $\ell_L$ ($\ell_R$) is the left (right) cavity inductance per unit length, $x_c$ is the position of the dividing capacitor $C_c$ in the central conducting strip of the cavity and $x_{\mathrm{cav}}$ is the length of the cavity. Including either the dividing capacitor in the CPW cavity or a resonator-resonator coupling capacitor of equal capacitance into our system produces matching low-energy spectra. An example of tuning the spectrum by manipulating the left cavity inductance per unit length is given in Fig.~\ref{fig:spectrum}.
For the cavity parameters used in Fig.~\ref{fig:spectrum}, the bare cavity resonance $\omega_L^b = \omega_R^b$ is achieved at $\ell_L = \ell_R$ [see Appendix~\ref{app:baremapping}]. Away from resonance, the spectrum is that of isolated bare cavities. Near resonance, the capacitive coupling of the cavity regions induces an avoided crossing in the spectrum near which the bare cavity eigenstates are mixed. The minimum energy gap $\Delta \epsilon^{\mathrm{min}}$ has a value of $\Delta \epsilon^{\mathrm{min}}/(2\pi\hbar)  = 32.82 \textrm{ MHz} $.

\begin{figure}[h]
\includegraphics[width=0.5\textwidth]{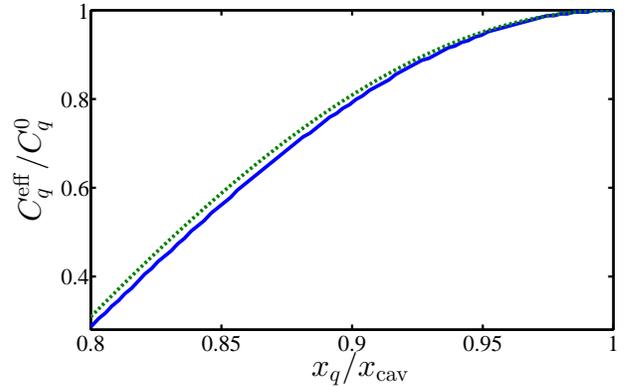}
\caption{(Color online) Effective resonator-qubit coupling capacitance in the LE model with respect to the position of the qubit in the right cavity region in the DE model chosen such that the low-energy excitation spectra of the two models are equal. The solid line shows the numerical solution obtained through matching the low-energy spectra of the LE and DE models. The dashed line gives the approximative effective capacitance $C_q^{\textrm{eff}} = C_q^0 \cos [\pi (x_q-x_{\textrm{cav}})/(x_{\textrm{cav}}-x_c)]$. We use the same parameters as in Fig.~\ref{fig:spectrum} with a fixed $\ell_L = \ell_R$. The cavity-qubit coupling capacitance in the DE model is $C_q^0=10\textrm{ fF}$, and the qubit excitation frequency is $\omega_q/(2\pi)= 12.62\textrm{ GHz}$.}
\label{fig:isolated}
\end{figure}

Due to the fact that we match the voltage operators at the ends of the cavity regions to our resonator voltage operators, placing the qubit at $x_q = x_{\textrm{cav}}$ inside the right cavity region will give matching spectra for the cavity-qubit system and our resonator-qubit system. Note that the DE model includes the qubit only after the divided cavity eigenmodes have been solved and, hence, the effective left capacitance $C_L$ in the LE model must not include $C_q$ in order for the two models to yield the same dynamics~\footnote{This is the method applied in this work. An alternative, more accurate choice is to include the effect of the coupling capacitance $C_q$ in both models. However, this will shift the frequencies only very slightly since typically $C_q$ is greatly smaller than the total cavity capacitance.}. If an offset in the qubit position is desirable, we may define an effective capacitance $C_q^{\textrm{eff}}$ in the LE model, selected such that the spectra of the DE and LE models still match. That is, we choose $C_q^{\textrm{eff}}$ such that the eigenproblems $(\hat{H}_{\mathrm{cav}} + \hat{H}_{c-q} + \hat{H}_q)\ket{\varphi} = \epsilon_{\varphi}\ket{\varphi}$, where $\hat{H}_{\mathrm{cav}}$ is the DE cavity Hamiltonian and $\hat{H}_{c-q}$ is the DE cavity-qubit coupling Hamiltonian, and $(\hat{H}_L + \hat{H}_R + \hat{H}_{L-R} + \hat{H}_q + \hat{H}_{R-q})\ket{\varphi'} = \epsilon_{\varphi'} \ket{\varphi'}$ give solutions for which $\{ \epsilon_{\varphi} \} = \{ \epsilon_{\varphi'} \}$. Figure \ref{fig:isolated} presents the dependence of $C_q^{\textrm{eff}}$ on the position of the qubit in the right cavity region.
For $x_q \approx x_{\textrm{cav}}$, the effective coupling capacitance may be approximated by $C_q^{\textrm{eff}} = C_q^0 \cos [\pi (x_q-x_{\textrm{cav}})/(x_{\textrm{cav}}-x_c)]$, where $C_q^0$ is the DE model cavity-qubit coupling capacitance~\footnote{The mapping procedure in Appendix~\ref{app:baremapping} matches the bare voltage operators at the ends of the cavity regions to the resonator voltage operators in our model. For $x_q \approx x_{\mathrm{cav}}$, we can approximate the cavity-qubit coupling by adding the spatial dependence of the bare cavity voltage operator in Eq.~(\ref{eq:VRb}) to the resonator-qubit coupling capacitance. This yields the approximate formula $C_q^{\textrm{eff}} = C_q^0 \cos [\pi (x_q-x_{\textrm{cav}})/(x_{\textrm{cav}}-x_c)]$. As the coupling is to the actual cavity modes, the approximation becomes less accurate when the qubit is positioned further from the end of the cavity.}. As shown in Fig.~\ref{fig:isolated}, the approximation is accurate in the vicinity of the exact matching point enabling one to exploit the simpler LE model even if the qubit is not placed at the very end of the right cavity region.

\section{Conclusions} \label{sec:conclusions}

We have presented and analyzed a scheme for a superconducting qubit coupled to a bosonic thermal bath through a sequence of harmonic oscillator modes. The specific device studied in detail is based on a pair of coupled quantum LC oscillators intermediating the interaction between the qubit and the bath. We have shown that the qubit-bath coupling can be controlled by manipulating the inductance of one of the resonators allowing for both rapid initialization to the ground state and protection from noise. The strong coupling to a low-temperature artificial environment means that we can lower the effective qubit temperature making the initialization of the qubit state more efficient.

Our design allows the use of lumped-element analysis rendering the analytical results more appealing than previous descriptions~\cite{scirep3.1987}. It grants us access to simple analytic solutions in several experimentally interesting regimes. Furthermore, the setup we propose here makes use of a capacitive coupling to the thermal noise source, a feature which can be easier to fabricate than the systems in the previous proposals~\cite{CavRevI,CavRevII,scirep3.1987} based on galvanic contacts. We propose a mapping between our scheme and a cavity-qubit design~\cite{scirep3.1987} and find that the results from the two scenarios are in good agreement. Our model works as the proof of principle for obtaining tunability in qubit-bath interactions by implementing a coupling through a sequence of resonators. Even though our analysis concentrated on a pair of LC resonators, we expect that a more general scheme in which the resistor is coupled to the qubit through $N$ resonators offers increased tunability. Especially, if all the $N$ resonators are out of resonance with each other and the qubit, we are likely to achieve $N$-fold protection as opposed to the two-fold protection observed here in Eq.~(\ref{eq:weakrate3}). However, the complexity of the experimental setup required to tune the resonance frequencies of multiple oscillators would consequently increase. Scaling up the number of qubits implemented using our setup would, in turn, result in linear increase in complexity as each qubit-bath coupling is individually controlled. Finally, the analysis presented in this paper is not only directly usable for the experimental realization of such tunability but also for devising more elaborate designs based on similar ideas and applications of these in various different physical systems. One promising candidate for such a design would be to extend the Purcell filter approach~\cite{Reed10} to our setup providing a possible direction of future studies.

\begin{acknowledgments}
We acknowledge the V\"ais\"al\"a Foundation and the Finnish National Doctoral Programme in Materials Physics (NGSMP) for financial support. This research has been supported by the Academy of Finland through its Centres of Excellence Program under Grant No. 251748 (COMP) in addition to the Grants No. 138903 and No. 135794. We have received funding from the European Research Council under Starting Independent Researcher Grant No. 278117 (SINGLEOUT).
\end{acknowledgments}

\appendix

\section{Analytical solution to the energy eigenproblem} \label{app:analyticalsolution}

In the low-energy regime and within the rotating-wave approximation, the time-independent Schr\"odinger equation $(\hat{H}_L + \hat{H}_R + \hat{H}_{L-R} + \hat{H}_q + \hat{H}_{R-q})\ket{\varphi_k} = \epsilon_k \ket{\varphi_k}$, where $k \in \{0,1,2,3\}$, has a ground state given by $\ket{\varphi_0} = \ket{0,0,g}$. See Eqs.~(\ref{eq:Hk})--(\ref{eq:Hresqb}) for definitions of the Hamiltonians. We select the zero-point energy such that $\epsilon_0 = 0$. After rewriting the above Hamiltonian in the basis described in Sec.~\ref{subsec:cqs}, the eigenproblem reduces to solving three simultaneous equations
\begin{align}
&\tan(\gamma_k) = \frac{\epsilon'_k-\omega_q}{g}, \label{eq:an1} \\
&\tan(\theta_k) = \frac{\epsilon'_k-\omega_q}{\epsilon'_k-\omega_L} \frac{-i\alpha}{\sqrt{|g|^2-(\epsilon'_k-\omega_q)^2}}, \label{eq:an2} \\
&\sin(\gamma_k) = \frac{-i(\epsilon'_k-\omega_q)}{\sqrt{|g|^2-(\epsilon'_k-\omega_q)^2}}, \label{eq:an3}
\end{align}
where $\epsilon'_k = \epsilon_k/\hbar$. The equations yield the eigenenergies $\epsilon_k$, $k = 1,2,3$, and, similarly, solving $\gamma_k$ and $\theta_k$ produces the excited states as $\ket{\varphi_k} = \sin(\theta_k) \ket{1,0,g} + \cos(\theta_k)\sin(\gamma_k)\ket{0,1,g} + \cos(\theta_k)\cos(\gamma_k)\ket{0,0,e}$.

\section{Weak-coupling limit} \label{app:weak}

Let us rewrite the total Hamiltonian of the resonator-qubit system [see Eqs.~(\ref{eq:Hk})--(\ref{eq:Hresqb}) for definitions] ignoring the resistor as $\hat{H}_{\mathrm{sys}} = \hat{H}_L + \hat{H}_R + \hat{H}_q + \hat{H}'$, where $\hat{H}' = \hat{H}_{L-R} + \hat{H}_{R-q}$ is the small perturbation in the weak coupling limit $|\alpha|, |g| \ll \omega_L,\omega_R,\omega_q$ and the system is far away from resonance $|\alpha|, |g| \ll |\omega_{l}-\omega_{s}|$, for all $l,s \in \{L,R,q\}$. The ground state is unaffected by the perturbation and given by $\ket{\varphi_0} = \ket{0,0,g}$ with its eigenenergy selected as the zero-point of energy. After applying the second-order perturbation theory in the small coupling parameters, the eigenenergies of $\hat{H}_{\mathrm{sys}}$ are given by 
\begin{align}
&\epsilon_{1} =\hbar \omega_L+\frac{\hbar \alpha^{2}}{\Delta_{RL}}, \label{eq:weaken1} \\
&\epsilon_{2} =\hbar \omega_R+\frac{\hbar \alpha^{2}}{\Delta_{LR}}+\frac{\hbar |g|^{2}}{\Delta_{qR}}, \\
&\epsilon_{3} = \hbar \omega_q+ \frac{\hbar |g|^{2}}{\Delta_{Rq}},
\end{align}
where we adopted the notation $\Delta_{ls} = \omega_l-\omega_s$ for the difference between two intrinsic system frequencies. The corresponding eigenstates are
\begin{widetext}
\begin{align}
& \ket{\varphi_1} = A_1\left[\left(\frac{2\Delta_{RL}^{2}-\alpha^{2}}{2\Delta_{RL}^{2}}\right) \ket{1,0,g} + \frac{\alpha}{\Delta_{RL}} \ket{1,0,g}+\frac{\alpha g}{\Delta_{RL}\Delta_{qL}}\ket{0,0,e} \right], \\
& \ket{\varphi_2}= A_2 \left[ \frac{\alpha}{\Delta_{LR}} \ket{1,0,g} + \left(\frac{2\Delta_{qR}^{2}\Delta_{LR}^{2}-|g|^{2}\Delta_{LR}^{2}-\alpha^{2}\Delta_{qR}^{2}}{2\Delta_{qR}^{2}\Delta_{LR}^{2}} \right) \ket{0,1,g} +\frac{g}{\Delta_{qR}}\ket{0,0,e}  \right], \\
& \ket{\varphi_3}=A_3\left[ \frac{g^*\alpha}{\Delta_{Rq}\Delta_{Lq}}\ket{1,0,g} + \frac{g^*}{\Delta_{Rq}}\ket{0,1,g} + \left(\frac{2\Delta_{Rq}^{2}-|g|^{2}}{2\Delta_{Rq}^{2}}\right)\ket{0,0,e} \right],
\end{align}
\end{widetext}
and the normalization constants are given by
\begin{widetext}
\begin{align}
A_1 &= \frac{2\Delta_{LR}^{2}\Delta_{Lq}}{\sqrt{4\alpha^{2}|g|^{2}\Delta_{LR}^{2}+\Delta_{Lq}^{2}(\alpha^{4}+4\Delta_{LR}^{4})}}, \label{eq:weakN1} \\
A_2 &= \frac{2\Delta_{LR}^{2}\Delta_{Rq}^{2}}{\sqrt{|g|^4\Delta_{LR}^4+2\alpha^{2}|g|^{2}\Delta_{LR}^{2}\Delta_{Rq}^{2}+\Delta_{Rq}^{4}\left(\alpha^{4}+4\Delta_{LR}^{4} \right)}}, \\
A_3 &= \frac{2\Delta_{qR}^{2}\Delta_{qL}}{\sqrt{4\Delta_{Rq}^{2}|g|^{2}\alpha^{2}+4\Delta_{Rq}^{4}\Delta_{Lq}^{2}+\Delta_{Lq}^{2}|g|^{4}}}. \label{eq:weakN3}
\end{align}
\end{widetext}

\section{Bare mode mapping} \label{app:baremapping}

If we omit the coupling capacitor inserted into the CPW cavity shown in Fig.~\ref{fig:cavity}, the cavity regions can be treated as individual resonators. We denote these resonators as \emph{bare} cavities and write the excitation frequencies for their lowest-energy modes as~\cite{Pozar04}
\begin{align}
\omega_L^b &= \frac{\pi}{x_c\sqrt{\ell_L c}}, \\
\omega_R^b &= \frac{\pi}{(x_\mathrm{cav}-x_c)\sqrt{\ell_R c}},
\end{align}
where $c$ is the constant cavity capacitance per unit length, $\ell_L$ ($\ell_R$) is the left (right) cavity inductance per unit length, $x_c$ is the position of the dividing capacitor $C_c$ in the central conducting strip of the cavity and $x_{\mathrm{cav}}$ is the length of the cavity. Similarly, the voltage operators for these modes are given by
\begin{align}
\hat{V}_L^b (x) &= V_L^{b0} \cos \left( \frac{\pi x}{x_c} \right) [(\hat{a}_L^b)^{\dagger}+\hat{a}_L^b], \\
\hat{V}_R^b (x) &= V_R^{b0} \cos \left[ \frac{\pi (x-x_{\mathrm{cav}})}{x_{\mathrm{cav}}-x_c} \right] [(\hat{a}_R^b)^{\dagger}+\hat{a}_R^b], \label{eq:VRb}
\end{align}
where $(\hat{a}_{l}^b)^{\dagger}$ and $\hat{a}_{l}^b$, $l=L,R$, are the creation and annihilation operators for the modes, respectively. The prefactors for the operators are $V_L^{b0} = \sqrt{\hbar \omega_L^b/(x_c c)}$ and $V_R^{b0} = \sqrt{\hbar \omega_R^b/[(x_\mathrm{cav}-x_c) c]}$.

We relate the bare cavities to our resonator system in Fig.~\ref{fig:setup} by demanding that $\omega_L = \omega_L^b$, $\omega_R = \omega_R^b$, $V_L^{b0} = V_L^0$ and $V_R^{b0} = V_R^0$. This causes the resonators in our system and the bare cavities to have equal excitation energies, and the prefactors of the voltage operators at the ends of the bare cavities to match the prefactors of the voltage operators in our system. The resulting mapping is given in Eqs.~(\ref{eq:bare1})--(\ref{eq:bare4}). Matching the voltage operators in this manner ensures that including the same coupling capacitors into both systems has the same effect on the spectrum. This is because the bare mode voltages at the coupling capacitor effectively determine its charging Hamiltonian.

\bibliography{draftbib}

\end{document}